\documentclass[twocolumn,english,11pt]{article}
\usepackage[intlimits,sumlimits]{amsmath}
\usepackage{amssymb,amsfonts}
\usepackage{graphicx}
\usepackage{hyperref}

\begin{document}
\title{Scanned gate microscopy of a one-dimensional quantum dot}

\author{Lingfeng M. Zhang and Michael M. Fogler\\
{\it Department of Physics, University of California San Diego,
9500 Gilman Drive,}\\
{\it La Jolla, California 92093}}

\date{\today}

\begin{abstract}

We analyze electrostatic interaction between a sharp conducting tip and
a thin one-dimensional wire, e.g., a carbon nanotube, in a scanned gate
microscopy (SGM) experiment. The problem is analytically tractable if
the wire resides on a thin dielectric substrate above a metallic
backgate. The characteristic spatial scale of the electrostatic coupling
to the tip is equal to its height above the substrate. Numerical
simulations indicate that imaging of individual electrons by SGM is
possible once the mean electron separation exceeds this scale
(typically, a few tens of nm). Differences between weakly and strongly
invasive SGM regimes are pointed out.

\end{abstract}

\maketitle


Scanned gate microscopy (SGM) is a recent addition to the arsenal of
modern scanned probe techniques that enable one to study and manipulate
nanoscale objects at a single-electron level. Much interest has been
attracted by experiments~\cite{Tans_00, Bachtold_00, Bockrath_00,
Woodside_02, Freitag_02, Staii_05} where the SGM has been employed for
probing carbon nanotube (CNT) quantum dots under the conditions of
Coulomb blockade. In those experiments, a movable sharp tip with a
controllable electrostatic potential was employed to add or remove
electrons from the dots one by one, which was monitored via transport
measurements.

\begin{figure}
\includegraphics[width=3in,clip=true]{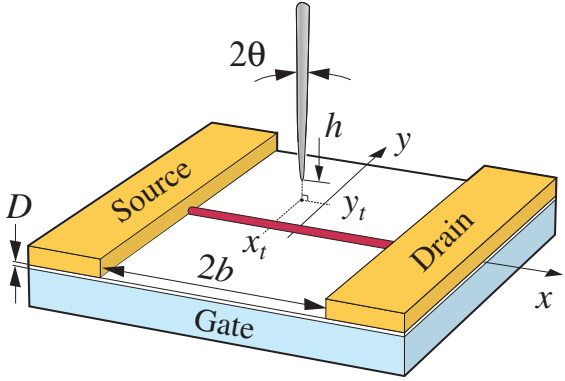}
\caption{(Color online) The SGM setup studied in this paper: a movable
metallic tip that probes a 1D wire (the horizontal cylinder).
\label{fig:exp setup}}
\end{figure}

The goal of this work is to explore the capabilities of the SGM for
imaging the real-space electron structure in a CNT dot. In parallel,
we develop an analytical approach that clarifies what controls the
fundamental limits of the SGM spatial resolution. We focus on a typical
SGM geometry where the CNT resides on a dielectric substrate above a
metallic backgate (Fig.~\ref{fig:exp setup}). We show below that if the
substrate is thin, so that the gate efficiently screens Coulomb
interactions, the requisite electrostatic problem is tractable. We find
the potential created by the charged tip in terms of elementary
functions. For realistic experimental parameters our formula proves to
be accurate to about 5\%, which enables us to test with confidence the
assumptions made about the same quantity in prior
literature~\cite{Woodside_02, Freitag_02, Meunier_04}. In particular, we
show that for the case of a sharp tip the spatial resolution of the SGM
is set by the tip's height above the substrate. For thick substrates,
the resolution degrades and becomes of the order of the geometric mean
of the tip height and the CNT length. Typical experimental parameters
put one in between these two limiting cases, with the net resolution of
the order of a few tens of $\text{nm}$. We propose that this may be
sufficient for imaging of individual charges in a few-electron quantum
dots and support this idea by numerical simulations. Our method of
analysis can be extended to other one-dimensional (1D) and quasi-1D
systems (multiple quantum dots, CNT networks, {\it etc\/}.), and so we
hope it may be a useful tool in the experimental practice of SGM.

%
%
%


We consider a CNT of radius $R$ resting on an insulating substrate of
thickness $D$ deposited on a metallic gate (Fig.~\ref{fig:exp setup}).
We assume that the contact resistances are high, so that the CNT behaves
as a quantum dot. Such a dot is probed by a movable conducting tip,
hovering at the height $h$ above the point $\textbf{r}_t = (x_t, y_t)$
of the substrate plane. The radius $w(\zeta)$ of the tip as a function
of distance $\zeta = z - h$ from its end is given by $w(\zeta) = \theta
\zeta$ at $l \ll \zeta \ll L$, where $\theta \ll 1$ is the apex angle
and $L$ is the full length of the tip. Parameter $l \ll h$ accounts for
the rounding of the tip's end~\cite{Comment_on_tip}.

Our first task is to analyze Coulomb interactions among the charges on
the CNT, gate, and the tip. For simplicity, we set all dielectric
constants to unity~\cite{Comment_on_epsilon}. We start with a
qualitative picture. The excess charge density $\rho_d(x)$ induced on
the CNT by the external gates can be deduced from classical
electrostatics of a thin metallic wire. Earlier microscopic
calculations~\cite{Meunier_04, Pomorskii_04, Paillet_05, Svizhenko_05}
proved that this is a good approximation when lengthscales of interest
exceed $e / \rho_d$. If the distance $D$ to the screening gate is much
smaller than the length of the CNT, $\rho_d(x)$ is nearly uniform. This
is because the effective interaction potential
\begin{equation}
  \begin{split}
    U(x) = \frac{e^2}{|x| + R} - \frac{e^2}{\sqrt{x^2 + 4 D^2}}
  \end{split}
  \label{eq:U_dd}
\end{equation}
between electrons on the dot is short-range. For $R \ll D$ the
capacitance per unit length of the CNT is $c_d = 1 / 2
\ln\left(2D/R\right)$. Actually, even without the screening gate, i.e.,
at $D = \infty$, $\rho_d(x)$ is nearly uniform, except near the
ends~\cite{Jackson_00}, because the potential $U(x) \propto |x|^{-1}$ is
only marginally long-range in 1D.

Let us now discuss the linear charge density $\rho_t(z)$ of the tip.
Interestingly, it also becomes uniform~\cite{Belaidi_97} if $h$ and $D$
are both small and the tip is sharp, $\theta \ll 1$. In this case
$w(\zeta)$ is a slowly varying function; therefore, a segment of the tip
at height $z$ above the substrate is similar to a piece of thin wire of
radius $w(z - h)$. The distance to the screening gate for this segment
is $z + D$. A quick estimate of the capacitance of the tip per unit
length $c_t$ can be obtained from the formula for $c_d$ by replacing $R
\to w(z - h) \simeq (z - h) \theta$ and $D \to z + D$, i.e.,
\begin{equation}
  c_{t} \simeq \frac{1}{2 \ln[2 (z + D) / w(z - h)]}
  \simeq \frac{1}{2 \ln (2 / \theta)}.
  \label{eq:c_t}
\end{equation}
This result is confirmed by a formal perturbation theory that yields
$\rho_t(z)$ as a power series of the small parameter $1 / |\ln \theta|$,
similar to Ref.~\cite{Jackson_00}. In the leading order we obtain
($\zeta \equiv z - h$)
\begin{equation}
  \frac{1}{\rho_t} = \ln\left[
    \frac{4 (L - \zeta) \zeta}{L + \zeta + 2 h + 2 D}
    \frac{\zeta + 2 h + 2 D}{w^2(\zeta)}
  \right].
  \label{eq:rho_t}
\end{equation}
This formula applies at all $\zeta$ except very near the ends of the
tip, $\zeta = 0, L$. We tested it by solving the electrostatic problem
for the tip and the gate numerically.
%
%
An excellent agreement was reached for, e.g., $\theta = 0.1 \approx
6^\circ$ and $h = 30\,\text{nm}$, which are not too difficult to achieve
experimentally~\cite{Comment_on_tip}. For such $\theta$ and $h$ the
nonuniformity of $\rho_t$ within the important region $0 < \zeta < 2 D$
(see below) is about 15\% for $D = 30$--$200\,\text{nm}$. These are the
parameters adoped in our subsequent simulations described shortly below.

\begin{figure}
\includegraphics[width = 2.8in,clip=true]{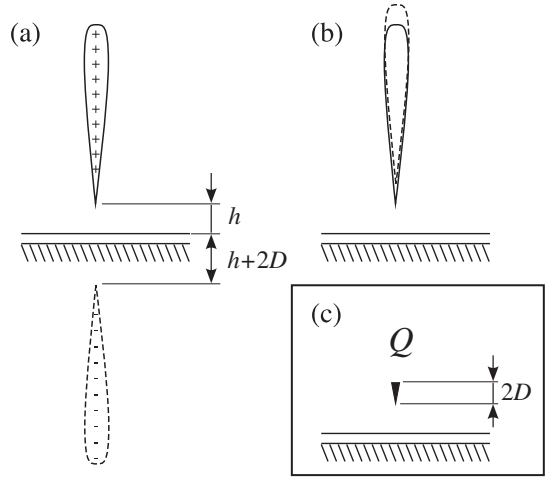}
\caption{Geometrical construction that enables one to replace a long
tip by a uniformly charged rod of a much smaller length $2 D$
(see main text).
\label{fig:tip_cartoon}}
\end{figure}

The approximate uniformity of $\rho_t$ entails a simple model for the
potential $U_t(\textbf{r})$ induced by the tip on the substrate. This
potential is created both by the tip itself and by its oppositely
charged image [Fig.~\ref{fig:tip_cartoon}(a)]. It is convenient to move
the latter into the $z > 0$ half-space by reflection
[Fig.~\ref{fig:tip_cartoon}(b)]. As a result, the two charge densities
cancel each other almost everywhere. The important uncompensated piece
is a uniformly charged rod of length $2D$
[Fig.~\ref{fig:tip_cartoon}(c)]. The potential induced by this rod is
\begin{equation}
  \begin{split}
    U_t = c_t V_t \left(
      \sinh^{-1}\frac{h+2D}{\Delta r}\right.
    \left.-\sinh^{-1}\frac{h}{\Delta r} \right),
  \end{split}
  \label{eq:U_t}
\end{equation}
where $\Delta r = |\textbf{r} - \textbf{r}_t|$ is the in-plane distance
from the tip and $V_{t}$ is the tips's voltage. In particular,
for a thin dielectric substrate, $U_t$ has the Coulomb form,
\begin{equation}
  U_t(\textbf{r})
  = {Q} / {\sqrt{\Delta r^2 + h^2}},\quad D \ll h.
  \label{eq:U_t_small_D}
\end{equation}
In this limit the effect of tip reduces to that of a point charge $Q = D
V_t / \ln (2 / \theta)$ positioned at height $h$ above the substrate
[Fig.~\ref{fig:tip_cartoon}(c)]. Clearly, $h$ sets the range of the
tip-dot interaction and is thus the sole geometric parameter that
controls the SGM spatial resolution ($\text{Res}$).

%
%

As $D$ increases, the resolution degrades. The following
estimate can be derived:
\begin{equation}
\text{Res} \sim \min\{
h + D,\ \sqrt{b h}
\},
\label{eq:Res}
\end{equation}
where $b \gg h$ is the half-length of the quantum dot probed by the
SGM. The reason why $b$ enters at large $D$ is the
slow decay of $U_t$ with $\Delta r$. In this case $\text{Res}$ is
determined not by the behavior of $U_t$ itself but by that of
its variation across the system, $\Delta
U_t(\textbf{r}) = U_t(\textbf{r}) - U_t(b, 0)$. Indeed, an additive
constant in the dot-tip interaction energy has no effect on spatial
resolution. A rough estimate of $\text{Res}$ is then
the value of $\Delta r$ at which $\Delta U_t$ drops to a half of its
maximum value at $\Delta r = 0$. Thus, in the case of a thick
substrate, $D \gg b$, where Eq.~(\ref{eq:U_t}) yields
\begin{equation}
  \Delta U_t \simeq c_t V_t \ln\left(
  \frac{b}{\sqrt{\Delta r^2 + h^2} + h}\right),\quad r_t \ll b,
  \label{eq:U_t_large_D}
\end{equation}
the second expression in Eq.~(\ref{eq:Res}) is recovered.

Note that the actual resolution of the SGM depends on temperature. At
low $T$ Coulomb blockade spectroscopy enables one to detect very small
variations of the energy of the system as a function of $\textbf{r}_t$.
Hence, the spatial resolution also improves. Nevertheless, outperforming
the estimate~(\ref{eq:Res}) by more than a factor of two or so would
probably require impractically low $T$.

The substrates used in the current SGM experiments commonly have the
\textit{effective\/} thickness of $D \sim
50\,\text{nm}$~\cite{Comment_on_epsilon}, which is not much larger than
a typical $h$. Therefore, Eq.~(\ref{eq:U_t_small_D}) is acceptable as a
first approximation. This also explains why an analogous model for the
tip --- a small metallic sphere --- postulated
previously~\cite{Freitag_02, Meunier_04} was in a reasonable agreement
with the experiment.

Armed with Eq.~(\ref{eq:U_t}), we now turn to the computation of
representative SGM patterns. The conductance measured in an SGM
experiment is directly related to the tip-dot capacitance $C_{dt}$. When
the preferred dot charge $Q_d = C_{dt} V_{t} + \text{const}$ is a
half-integer multiple of $e$, the Coulomb blockade is lifted and the
conductance peak is measured~\cite{Ingold_92}. The problem reduces to
calculation of $C_{dt}(\textbf{r}_t)$. Let us first discuss the case of
a metallic CNT where it can be done analytically.

The key idea is to treat the inverse self-capacitances of the dot and
the tip, $\ln (2 D / R)$ and $\ln (2 / \theta)$, respectively, as large
parameters. Then $C_{dt}$ can be computed by a perturbation theory. In
the leading order one needs to integrate the product $U_t(x) \rho_d(x)$
over the length of the dot. This gives the tip-dot interaction energy
from which $C_{dt}$ can be extracted. In the simplest case $D \ll h$ one
can use Eq.~(\ref{eq:U_t_small_D}) to get
\begin{equation}
  C_{dt} \simeq -\frac{
    D\ln\left(\frac{
        x_t+b+\sqrt{\left(x_t+b\right)^2+y_t^2+h^2}
      }{
        x_t-b+\sqrt{\left(x_t-b\right)^2+y_t^2+h^2}
      }\right)
  }{
    2\ln(2 / \theta)\ln(2 D / R)
  }
  \label{eq:C_dt}
\end{equation}
In deriving this equation, additional screening by source and drain
leads was neglected. This is legitimate away from the leads, at $b -
|x_t| \gg D, h_l$, where $h_l$ is the height by which the leads
rise above the substrate.

\begin{figure}
\includegraphics[width=2.8in,clip=true]{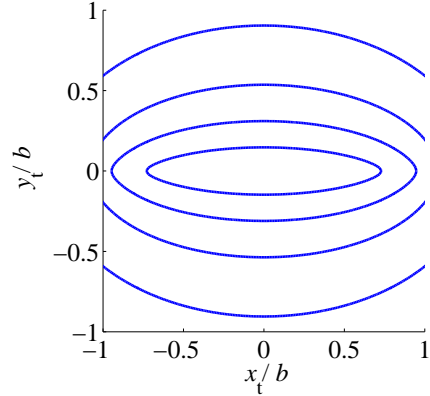}
\caption{High-conductance contours $C_{dt} V_t = (N + 1/2) e =
\text{const}$ as function of the tip position for a metallic CNT. $N$
runs through a set of consequtive integers. Parameters: $b =
250\,\text{nm}$, $R = 2\,\text{nm}$, $D = 30\,\text{nm}$, and $V_{t}
= 0.3\,\text{V}$.
\label{fig:Cdt_contour}}
\end{figure}

The contours of $Q_d(\textbf{r}_t) = C_{dt} V_t = (N + 1/2) e =
\text{const}$ computed according to Eq.~(\ref{eq:C_dt}) for typical
experimental parameters are presented in Fig.~\ref{fig:Cdt_contour}.
They are oval-shaped near the CNT and become nearly circular at larger
$N$, as in experiment~\cite{Woodside_02}. To verify our results for
$C_{dt}$ quantitatively, we compared them with the output of the
numerically exact capacitance calculator {FASTCAP}~\cite{FASTCAP}. For
parameters specified in Fig.~\ref{fig:Cdt_contour} and the tip of length
$L = 10\,\mu\text{m}$ we found a 5\% agreement, which is gratifying
given the simplicity of our approach and the approximations made.

%
%

Classical electrostatics we used thus far is valid for a
\textit{metallic\/} nanotube where the average inter-electron separation
$a$ is small. A more interesting question is whether the SGM can help us
to see effects beyond continuum electrostatics, for example: {\it Can
SGM image individual electrons?\/}

\begin{figure}
\includegraphics[width=2.9in,clip=true]{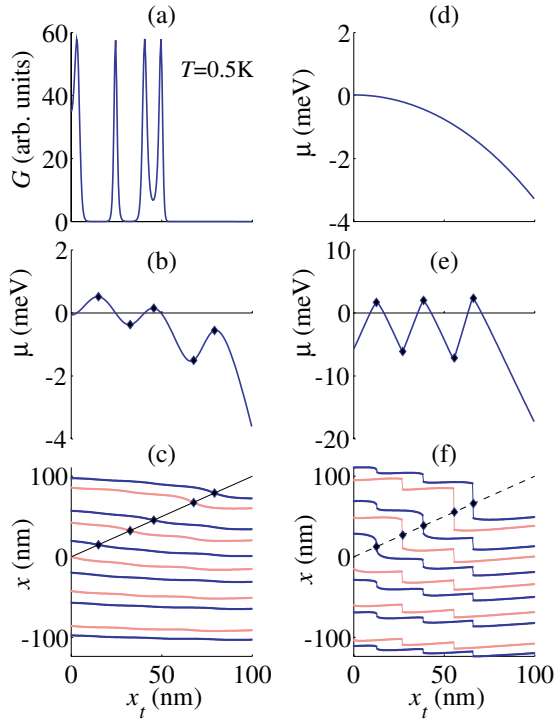}
\caption{(Color online) Results of numerical simulations for SGM of
Wigner molecules. (a) Conductance [estimated per
Eq.~(\ref{eq:G_orthodox_single})] (b) chemical potential, and (c)
ground-state electron positions for $h = 30\,\text{nm}$, $D = 200\,{\rm
nm}$, and $V_{t} = -0.1\,\text{V}$. In plot (c) every other curve from
the top is for the six-electron molecule, the rest are for the
five-electron state. At a higher tip voltage, $V_{t} = -0.25\,\text{V}$,
plot (c) evolves into (f) and plot (b) into (e). At a larger tip height,
$h = 90\,\text{nm}$, graph (b) evolves into (d). Diamonds correspond
to $x_t$ at which the tip is directly above one of the electrons.
\label{fig:G_mu_xe}}
\end{figure}

As explained earlier, $\text{Res}$ is of the order of the height $h$ of
the tip above the substrate, typically, a few tens of nm. This is
insufficient for resolving individual electrons in a metallic CNT. On
the other hand, in a semiconducting CNT the mean inter-particle
separation $a = 20$--$30\,\text{nm}$ can be achieved. In such CNTs
electrons (or holes) can be added to the conduction (valence) band one
by one starting from zero~\cite{Jarillo-Herrero_04}. Since the typical
size of a CNT quantum dot is a fraction of a micron, it would contain
only a few charge carriers altogether. In this regime the electrons form
a Wigner molecule: a state where they are highly localized near the
classical equilibrium positions. The Wigner-molecule approximation is
accurate if $r_s \equiv a / 2 a_B > 4$,~\cite{Reimann_02, Fogler_06}
where $a_B$ is the effective Bohr radius. At large $r_s$ the charge
distribution of each electron can be approximated by a Gaussian with the
standard deviation $w \approx 0.52 a_B (a / a_B)^{3 / 4}$. For example,
for $a \approx 40\,\text{nm}$ (as in Fig.~\ref{fig:G_mu_xe}) and $a_B
\sim 1.4\,\text{nm}$ (as in the experiment~\cite{Jarillo-Herrero_04}) we
get $w \approx 6.4\,a_B$, in agreement with numerical simulations of
Wigner molecules~\cite{Reimann_02}.

Suppose $\text{Res} < a$, then as the tip moves along the dot, it
interacts primarily with the nearest electron, and so continuum
electrostatics is no longer valid. Instead, one expects a significant
oscillatory modulation of the tip-dot coupling: maxima when the
(repulsive) tip is positioned directly above an electron and minima when
it is between two adjacent electrons. Such variations may be detectable
by examining the spacing of the conductance peaks.

%
%

To verify these qualitative ideas we carried out a series of numerical
simulations. The Hamiltonian of the system $H$
was taken to be
%
\begin{align}
&H = \sum_{1\leq i < j \leq N} U(x_i - x_j) + \sum_{i} e U_t(x_i)
   + H_{\text{con}},\notag\\
&H_{\text{con}} = \sum_{i = 1}^{N} e [\phi_{\text{con}}(x_i) + V_g],
\end{align}
%
where electron-electron and electron-tip interactions are given by
Eqs.~(\ref{eq:U_dd}) and (\ref{eq:U_t}), $V_g$ is the gate voltage
relative to the that of the leads, and $\phi_{\text{con}}(x)$ is the
confinement potential due to contact potential difference between the
leads and the CNT. Following Ref.~\cite{Jarillo-Herrero_04}, we chose
$\phi_{\text{con}}(x) = (k/2) (x^2 - b^2)$, with the
coefficient $k$ close to the number estimated in that paper. The
ground-state energy $E_N$ and the electron positions $x_j$ were found
numerically for a dense grid of tip coordinates on the interval $0 < x_t
< b$. Figure~\ref{fig:G_mu_xe} shows the results. Unlike
Fig.~\ref{fig:Cdt_contour} here we focus on line scans where the tip
stays directly above the CNT, i.e., at $y_t = 0$.

At low $T$ there are only two important charge states to consider, $Q_d
= N e$ and $Q_d = (N + 1)e$. In the simulations the gate voltage $V_g$
was adjusted to have $N = 5$. The dependence of the electron positions
on $x_t$ in the two competing configurations is shown in
Fig.~\ref{fig:G_mu_xe}(c) for the relatively low voltage of $V_t =
-0.1\,\text{V}$. The displacements of the electrons by the tip are
rather small, so here the tip is a weakly invasive probe. Shown by
diamonds are the tip positions where it is directly above an electron.
The chemical potential $\mu \equiv E_{N + 1}-E_{N}$ for the same SGM
scan is plotted in Fig.~\ref{fig:G_mu_xe}(b). It exhibits the expected
spatial variations: maxima and minima near $x_t$'s of the diamonds. At
the intersection of the $\mu(x_t)$ curve with the horizontal line $\mu =
0$ the two charged states become degenerate. At such points the
conductance $G$ through the dot has peaks. The calculation of the shape
of these peaks is beyond the scope of this work. However, for
illustrative purposes, in Fig.~\ref{fig:G_mu_xe}(a), we plot the
expression
\begin{equation}
           G = \frac{1}{2 T R_{sd}} \frac{\mu}{\sinh (\mu / T)},
\label{eq:G_orthodox_single}
\end{equation}
which holds for large metallic dots~\cite{Ingold_92}. Parameter $R_{sd}$
here denotes the sum of the tunneling resistances at the source and the
drain leads.

According to the discussion above, imaging the electron positions
amounts to finding the maxima and minima of the chemical potential. To
achieve that in experiment one would need to repeat SGM scans at
different $V_g$. Let $\mu(x_t)$ be the chemical potential at some
initial gate voltage $V_g^{*}$. At a different $V_g$, the chemical
potential changes by a constant, i.e., the curve in
Fig.~\ref{fig:G_mu_xe}(b) shifts up or down as a whole. The conductance
peaks are found at the roots of the equation $\mu(x_t) = e V_g^{*} - e
V_g$. Based on this relation, the entire curve of $\mu(x_t)$ can be
deduced by tracking the peak positions as a function of $V_g$.

As the tip's voltage increases, it perturbs the system stronger, see
Fig.~\ref{fig:G_mu_xe}(f). Here we focus on the case of a repulsive tip,
which expels electrons from underneath itself. This creates a double
quantum dot: Dot 1 in front of the tip and Dot 2 behind it [regions
above and below the dashed line in Fig.~\ref{fig:G_mu_xe}(d),
respectively]. As the tip moves, the following sudden changes in the
double-dot system take place: (1) an electron is expelled from Dot 1
into the lead, which occurs when $\mu < 0$ changes to $\mu > 0$, (2)
another electron enters Dot 2 from the opposite lead, at the points
where $\mu > 0$ changes to $\mu < 0$, and (3) an electron jumps between
Dots 1 and 2, at the positions labelled by diamonds in
Fig.~\ref{fig:G_mu_xe}(f). Under the assumptions that contact resistance
$R_{sd}$ is higher than the inter-dot tunneling resistance and the
temperature is low enough, the peaks in total conductance $G$
coincide~\cite{Matveev_96} with events (1) and (2), i.e., the points
$\mu = 0$.

The strongly invasive SGM can be easily distinguished from a weakly
invasive one experimentally. First, in the invasive regime $\mu(x_t)$
curve has cusps [Fig.~\ref{fig:G_mu_xe}(e)] whereas in the weakly
invasive one it varies smoothly [Fig.~\ref{fig:G_mu_xe}(b)]. Second, in
the double-dot case the peak heights of the conductance have a strong
modulation (beating pattern)~\cite{Kouwenhoven_03} as a function of
$x_t$ and $V_g$. The highest $G$ is achieved if the Coulomb blockade in
the two dots is lifted simultaneously. This can be realized by tuning
$V_g$ to some special values at which cusps of $\mu(x_t)$ touch the $\mu
= 0$ line. We do not attempt to discuss $G$ in any more detail because
its calculation for the double-dot is a complicated
problem~\cite{Matveev_96}. Equation~(\ref{eq:G_orthodox_single}) becomes
a poor approximation in this regime.

The effect of the tip height on spatial resolution is demonstrated in
Fig.~\ref{fig:G_mu_xe}(d): for $h = 90\,\text{nm}$ the oscillations in
$\mu(x_t)$ are completely obliterated. This confirms that the resolution
limit strongly depends on $h$. In contrast, having a rather large $D =
200\,\text{nm}$ does not degrade the resolution much. It may be
surprising at first but, in fact, Eq.~(\ref{eq:Res}) gives $\text{Res}
\sim 55\,\text{nm}$ (using $b = \max|x_j| \approx 100\,\text{nm}$),
which is only slightly larger than the mean electron spacing $a \approx
40\,\text{nm}$. Note also that a rather low temperature, $T =
0.5\,\text{K}$, is chosen.

\begin{figure}
\includegraphics[width=2.8in,clip=true]{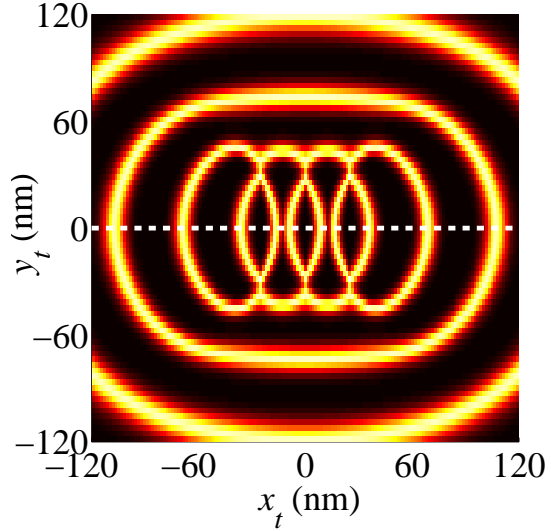}
\caption{(Color online) Contours of high conductance in a
low-$T$ two-dimensional SGM scan. Simulation parameters: $D = 30\,\text{nm}$,
$h = 30\,\text{nm}$, $V_{t} =-0.4\,\text{V}$, $T = 15\,\text{K}$.
\label{fig:G_contour}}
\end{figure}

Finally, let us discuss the results for a two-dimensional SGM scan,
Fig.~\ref{fig:G_contour}. We once again see contours of $G$, akin to
Fig.~\ref{fig:Cdt_contour} but with striking differences. One of the
contours contains pronounced oscillatory features --- ``wiggles''.
Another contour (closest to the CNT) has been broken into three
disconnected pieces, as though the amplitude of the corresponding
wiggles exceeded this contour's original width. Similar to the 1D scans
of Fig.~\ref{fig:G_mu_xe}, such effects originate from discreteness of
the electron charge that causes modulation of the chemical potential as
the tip passes by individual electrons. In general, the contour with $N$
wiggles separates the interior region of $N - 1$ electrons from the
exterior one with $N$ electrons in the ground-state. The lack of
wiggles on more distant contours is explained by noting that the spatial
resolution is now set by $\sqrt{h^2 + y_t^2}$, i.e., it worsens as the
tip is moved laterally away from the dot. At large $y_t$ continuum
electrostatics applies, and so the distant $G$-contours are similar
to those of Fig.~\ref{fig:Cdt_contour}.
 
%
%

To assess how stable the predicted pattern is against thermal
smearing, $G$ in Fig.~\ref{fig:G_contour} is computed for $T =
15\,\text{K}$. For simplicity, the calculation is done based on
Eq.~(\ref{eq:G_orthodox_single}) although this equation is a poor
approximation in this case. In particular, it gives a constant intensity
along the contours of high $G$ whereas in reality it is expected to have
strong variations. A more accurate calculation of $G$ is deferred for
future work.

In conclusion, we have studied electrostatics of a prototypical SGM
experiment in which a CNT quantum dot is probed by a sharp metallic tip.
We developed an analytical model for the coupling between the tip and
charges on the sample plane and verified it by numerical simulations.
Our model explains why the resolution is controlled primarily by the
tip-dot distance (the height above the substrate and the lateral
distance away from the CNT) rather than by much larger overall
dimensions of the tip and the dot. We predicted that the SGM of a
semiconducting CNT in the regime of strong electron-electron
interaction may be able to detect positions of individual electrons
forming a Wigner molecule. Lastly, the tip can act as either a
weakly or a strongly invasive probe, and so the SGM can be employed both
for single-electron imaging and for nano-manipulation.

This work is supported by UCSD ASCR and C.\&~W.~Hellman Fund. We thank
S.~Kalinin, P.~McEuen, and J.~Zhu for valuable comments.



\begin{thebibliography}{99}





\bibitem{Tans_00} S.~J.~Tans and C.~Dekker,
Nature (London) {\bf 404}, 834 (2000).

\bibitem{Bachtold_00} A.~Bachtold, M.~S.~Fuhrer, S.~Plyasunov,
M.~Forero, E.~H.~Anderson, A.~Zettl, and P.~L.~McEuen,
Phys.\ Rev.\ Lett.\ {\bf 84}, 6082 (2000).

\bibitem{Bockrath_00} M.~Bockrath, W.~Liang, D.~Bozovic, J.~H.~Hafner,
C.~M.~Lieber, M.~Tinkham, and H.~Park,
Science {\bf 291}, 283 (2000).

\bibitem{Woodside_02} M.~Woodside,
PhD thesis, University of California, Berkeley, 2001;
M.~T.~Woodside and P.~L.~McEuen,
Nature (London) {\bf 296}, 1098 (2002).

\bibitem{Freitag_02} M.~Freitag, S.~V.~Kalinin, D.~A.~Bonnell, and
A.~T.~Johnson,
Phys.\ Rev.\ Lett.\ \textbf{89}, 216801 (2002).

\bibitem{Staii_05} C.~Staii, A.~T.~Johnson, R.~Shao, and D.~A.~Bonnell,
NanoLett.\ \textbf{5}, 893 (2005).




\bibitem{Meunier_04} V.~Meunier, S.~V.~Kalinin, J.~Shin, A.~P.~Baddorf,
and R.~J.~Harrison,
Phys.\ Rev.\ Lett.\ \textbf{93}, 246801 (2004);

\bibitem{Comment_on_tip} Most commonly, $\theta \sim 0.25$ and $l \sim
20\,\text{nm}$ but much smaller values are feasible with ion-milled
tips.

\bibitem{Comment_on_epsilon} If the substrate has the dielectric
constant $\epsilon \neq 1$, its thickness is effectively
reduced by $\epsilon$, e.g., for the geometrical thickness of
$200\,\text{nm}$ and $\epsilon = 3.9$ of SiO$_2$, $D \approx
50\,\text{nm}$.

\bibitem{Pomorskii_04} P.~Pomorski, L.~Pastewka, Ch.~Roland,
H.~Guo, and J.~Wang,
Phys.\ Rev.\ B\ {\bf 69}, 115418 (2004).

\bibitem{Paillet_05} M.~Paillet, P.~Poncharal, and A.~Zahab,
Phys.\ Rev.\ Lett.\ \textbf{94}, 1868011 (2005), 
and references therein.

\bibitem{Svizhenko_05} A.~Svizhenko and M.~P.~Anantram,
Phys.\ Rev.\ B\ {\bf 72}, 85430 (2005).


\bibitem{Jackson_00} J.~D.~Jackson,
Am.\ J.\ Phys.\ {\bf 68}, 789 (2000).

\bibitem{Belaidi_97} S.~Belaidi, P.~Girard, and G.~Leveque,
J.\ Appl.\ Phys.\ {\bf 81}, 1023 (1997).

\bibitem{Ingold_92} \textit{Single-Charge Tunneling\/},
edited by H.~Grabert and M.~H.~Devoret (Plenum, New York, 1992).

\bibitem{FASTCAP} FastCap, Research Laboratory of Electronics, MIT.
Available at \texttt{http://www.rle.mit.edu}.

\bibitem{Jarillo-Herrero_04} P.~Jarillo-Herrero,
S.~Sapmaz, C.~Dekker, L.~P. Kouwenhoven, and H.~S.~J. van der Zant,
Nature (London) \textbf{429}, 389 (2004).

\bibitem{Reimann_02} For a review, see S.~M.~Reimann and M.~Manninen,
Rev.\ Mod.\ Phys.\ \textbf{74}, 1283 (2002).

\bibitem{Fogler_06} M.~M.~Fogler and E.~Pivovarov,
Phys.\ Rev.\ B\ \textbf{72}, 195344 (2005).

\bibitem{Matveev_96} K. A. Matveev, L. I. Glazman, and H. U. Baranger,
Phys.\ Rev.\ B\ {\bf 54}, 5637 (1996).

\bibitem{Kouwenhoven_03} W. G. van der Wiel, S. De Franceschi,
J. M. Elzerman, T. Fujisawa, S. Tarucha, and L. P. Kouwenhoven,
Rev.\ Mod.\ Phys.\ \textbf{75}, 1 (2003).


\end{thebibliography}
\end{document}